\documentclass[aip,jap,reprint]{revtex4-1}
\usepackage{graphicx}
\usepackage{dcolumn}
\usepackage{bm}
\usepackage{amsmath}
\usepackage{siunitx}
\usepackage{xcolor}

\begin{document}

\title{High-Field Q-slope Mitigation due to Impurity Profile in Superconducting Radio-Frequency Cavities} 
\author{M.~Checchin}
\email[]{checchin@fnal.gov}
\affiliation{Fermi National Accelerator Laboratory, Batavia IL 60510, USA}
\author{A.~Grassellino}
\affiliation{Fermi National Accelerator Laboratory, Batavia IL 60510, USA}

\date{\today}

\begin{abstract}
In this study, we present recent insights on the origin of the high-field Q-slope in superconducting radio-frequency cavities. Consequent hydrofluoric acid rinses are used to probe the radio-frequency performance as a function of the material removal of two superconducting bulk niobium cavities prepared with low temperature nitrogen infusion. The study reveals that nitrogen infusion affects only the first few tens of nanometers below the native oxide layer. The typical high-field Q-slope behavior of electropolished cavities is indeed completely recovered after a dozen hydrofluoric acid rinses. The reappearance of the high-field Q-slope as a function of material removal was modeled by means of London's local description of screening currents in the superconductor, returning good fitting of the experimental data and suggesting that a layer of interstitial impurities with diffusion length of the order of tens of nanometers can mitigate high-field Q-slope.    
\end{abstract}


\maketitle 

One of the phenomena still source of debate governing the behavior of superconducting radio-frequency (SRF) cavities is the so-called high-field Q-slope (HFQS). HFQS presents itself as the rapid drop of the quality factor as a function of the accelerating gradient for peak magnetic fields higher than 100~mT. This phenomenon is observed in cavities degassed at \SI{800}{\degreeCelsius} for 3 hours that received standard electropolishing\cite{Padamsee_Book2} (EP) or buffer chemical polishing\cite{Padamsee_Book2} (BCP).

The origin of the HFQS was initially attributed to several possible phenomena.\cite{Padamsee_Book2,Ciovati_PRSTAB_2010} Currently, the mechanism that better describes the experimental data involves the occurrence of nano-hydrides proximity-coupled\cite{deGennes_RevModPhys_1964,Orsay_PKM_1967} to the superconducting niobium matrix that allow for the abrupt increase of the surface resistance once the electromagnetic field in the cavity is raised above their breakdown condition, turning them normal-conducting.\cite{Romanenko_SUST_2013}

Historically, the process discovered able to mitigate HFQS was the mild baking at $\SI{120}{\degreeCelsius}$ for 48 hours \textit{in situ}.\cite{Visentin_EPAC_1998} Such a treatment, so-called $\SI{120}{\degreeCelsius}$ baking, allows for accelerating gradients higher than 35~MV/m (in TESLA-type accelerating structures at 1.3~GHz),\cite{TESLA_Cavities_PRST_2000_short} with quality factor (Q-factor) in the range $5\cdot10^9$ to $1\cdot10^{10}$. It was also observed that nitrogen doping\cite{Grassellino_SUST_2013} is mitigating effectively HFQS, though in many cases N-doped cavities do not reach the threshold field due to early quench.

In the framework of nano-hydrides precipitation, extensive work was performed to pinpoint the effect of the $\SI{120}{\degreeCelsius}$ baking in the nucleation and growth of such precipitates. In particular, generation of vacancies during the treatment\cite{Romanenko_APL_2013, Romanenko_APL_2014_1} and the presence of impurities\cite{Grassellino_SUST_2013,Ford_SUST_2013,Garg_SUST_2018} were suggested as possible ways of binding interstitial hydrogen preventing its precipitation into niobium hydrides upon cooling. It was also shown that the effect of the $\SI{120}{\degreeCelsius}$ baking resides in the initial tens of nanometers just below the native oxide,\cite{Ciovati_PRSTAB_2007,Romanenko_PRST_2013} where a disorder profile associated to vacancies nucleating during the treatment was found.\cite{Romanenko_APL_2014_1} 

In 2016, a novel treatment able to mitigate HFQS was discovered at Fermilab.\cite{Grassellino_SUST_2017} Such a thermal process, so-called nitrogen infusion (N-infusion), consists in treating the cavity with nitrogen at $\SI{120}{\degreeCelsius}$ for 48 hours after an initial degas at $\SI{800}{\degreeCelsius}$ for 3 hours. Cavities so prepared allowed for gradients of the order of 45~MV/m without HFQS, and Q-factor of the order of 10$^{10}$.
\begin{figure*}[t]
\centering
\includegraphics[scale=0.9]{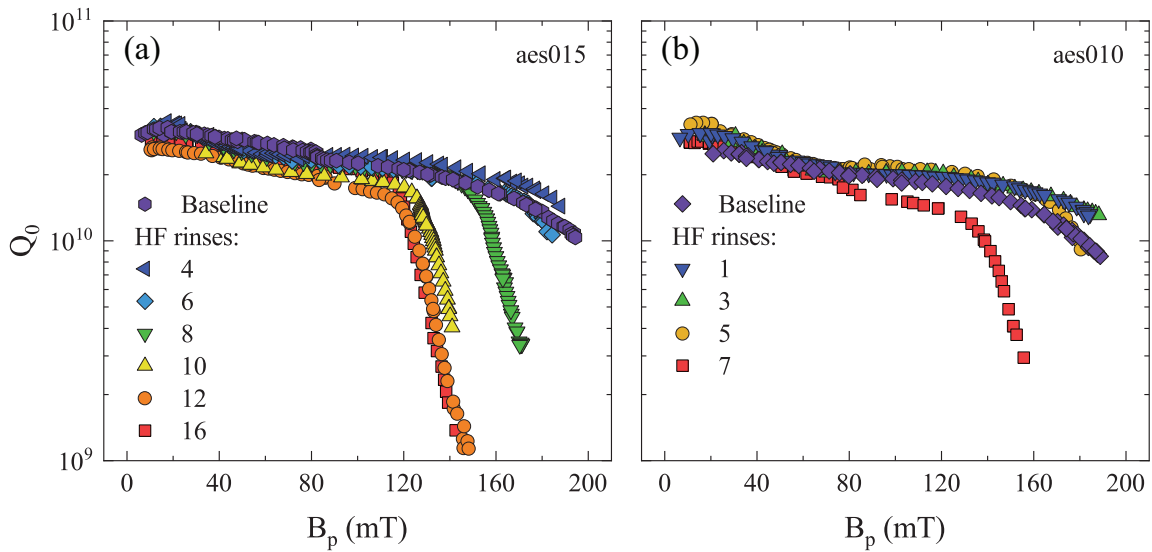}
\caption{The Q-factor versus peak magnetic field acquired at 2~K of the two cavities studied (aes015 in (a) and aes010 in (b)) is reported after subsequent material removals via HF rinsing.}
\label{fig.data}
\end{figure*}

In this Letter, we present recent insights on the N-infusion process, we demonstrate it affects only the first tens of nanometers of material and we discuss how it mitigates the HFQS. The conclusions of this work are of general interest for the SRF community and are of central importance for further improving the understanding of the HFQS in SRF cavities.

Two superconducting radio-frequency (SRF) single-cell TESLA-type\cite{TESLA_Cavities_PRST_2000_short} bulk niobium cavities (aes010 and aes015) where N-infused\cite{Grassellino_SUST_2017} at $\SI{120}{\degreeCelsius}$ with the following procedure. The cavities received bulk electropolishing ($\sim$150~$\mu$m EP) and were then baked at $\SI{800}{\degreeCelsius}$ for 3 hours to degas hydrogen. After this step the temperature was lowered to $\SI{120}{\degreeCelsius}$ and kept constant for 48 hours while nitrogen, with partial pressure of 25~mTorr, was inlet in the furnace. The cavities so prepared were high pressure water rinsed and RF-tested at the vertical test facility of Fermilab. The performance of the two cavities as a function of the peak magnetic field at the surface ($B_p$) are reported in the two graphs of Fig.~\ref{fig.data} and labeled with ``baseline''. Both cavities reached fields of the order of 45~MV/m ($\sim$190~mT, the conversion factor is reported in Ref.~\onlinecite{TESLA_Cavities_PRST_2000_short}).

Consequent tests were performed after removing material from the inner surface by means of hydrofluoric acid (HF) rinsing, similarly to what was done in Ref.~\onlinecite{Romanenko_PRST_2013}. Whenever HF is in contact with the cavity surface, it dissolves the native oxide (with a thickness of approximately $\sim$4-5~nm, as shown by high resolution TEM images)\cite{Trenikhina_JAP_2015}, and upon subsequent water rinse, a new oxide layer is grown, consuming about 2~nm of niobium. This estimation can be carried out by determining the number of Nb atoms in 4-5~nm of Nb$_2$O$_5$ and converting it in thickness of metallic niobium. The estimated thickness must be taken as the approximate maximum possible removal rate per HF rinse, since the time between multiple HF rinses in series might not be long enough to allow for the full-thickness growth of the oxide.

In both cavities, the HFQS onset could be observed after at least four HF rinses, which correspond to about 8~nm of removed material. HFQS was completely recovered after twelve HF rinses, when the Q-factor versus peak magnetic field ceased to evolve as a function of material removal. This finding suggests that the effect of nitrogen infusion resides in the first tens of nanometers of material, similarly to what was observed for the baking at \SI{120}{\degreeCelsius}.\cite{Ciovati_PRSTAB_2007,Romanenko_PRST_2013}

This finding is in good agreement with TOF-SIMS measurements of N-infused cut-outs, extracted from a cavity that received the same treatment as the cavities under study.\cite{Romanenko_SRF_2019} It was observed that the low temperature process at $\SI{120}{\degreeCelsius}$ allows nitrogen to enter the material and create a diffusion profile of about 15~nm. The HFQS disappearance after N-infusion is most likely linked to this enriched nitrogen layer just underneath the native oxide layer, since a layer of material of comparable thickness must be removed in order to observe HFQS.

It is important to point out that a minor leak of air in the furnace during the baking of aes010, aes015, and the cavity from which the cut-outs were extracted was later discovered. The air contamination is the most likely culprit for the lower Q-factor observed for these cavities in the ``baseline'' status compared to previous N-infused resonators.\cite{Grassellino_SUST_2017} A couple of HF rinses (specifically 4 HF rinses for aes015 and 1 HF rinse for aes010) were enough to remove the contamination and recover the Q-factor, especially at high field. The TOF-SIMS analysis\cite{Romanenko_SRF_2019} highlighted also an oxygen profile extending for about 100~nm in the material, most likely generated by the air leak as well. In this case, the presence of oxygen cannot be completely disregarded and could also contribute in the HFQS mitigation.

It is theoretically expected that, diffusion layers of interstitial impurities in niobium increase the penetration of the screening currents into cleaner regions with larger critical currents, thus limiting the suppression of the screening current to a thin ``dirty'' region close to the surface, allowing the superconductor to bear higher surface magnetic field.\cite{Wave_PRR_2019} The same current redistribution effect may play a role in the HFQS phenomenon, since impurities profiles seem to be involved in the mitigation of HFQS as well.

It is expected that hydrides precipitation is limited by the occurrence of vacancies\cite{Romanenko_APL_2013,Romanenko_APL_2014_1} and impurities,\cite{Grassellino_SUST_2013, Ford_SUST_2013,Garg_SUST_2018} in turn mitigating HFQS. In N-doped and \SI{120}{\degreeCelsius} baked cavities it was indeed observed a lower number of nano-hydrides\cite{Trenikhina_JAP_2015,Sung_LCWS_2018,Sung_SRF_2019} compared to clean EP cavities. However, these mechanisms, even if occurring, cannot be accounted as the only mitigation phenomena to eliminate HFQS. As observed via cryogenic atomic force microscopy measurements, nano-hydrides are forming at the cavity surface independently on the processing and are forming also after $\SI{120}{\degreeCelsius}$ baking or N-infusion.\cite{Sung_LCWS_2018, Sung_SRF_2019}

As we will describe in the following part of this Letter, the current redistribution due to impurity profiles can mitigate the HFQS, even if nano-hydrides are present at the surface. 

Whenever impurity profiles occurs in the material, the penetration depth of the magnetic field ($\lambda$) varies with depth and it can be shown\cite{Pambianchi_PRB_1994} that the perfect diamagnetic behavior in the London theory framework is described by
\begin{equation}
\lambda^{2}B''(x)+2\lambda\lambda'B'(x)-B(x)=0\text{.}
\label{eq.london}
\end{equation}

We expect that impurity diffusion in the material follows the constant-source diffusion law, typical for impurity diffusion in bulk materials. Accordingly, the penetration depth will follow the variation in impurity concentration, saturating to its constant ``clean'' value in the bulk\textemdash $\lambda_0=39$~nm.

On the other hand, we do not expect that in this specific instance nano-hydrides can play a role in modifying the mean-free-path in the material. The average distance between them was shown to be in the order of microns\cite{Sung_SRF_2019}, while typical values of mean-free-path in niobium with impurity profiles at the surface is on the order of tens to hundreds of nanometers\cite{Martinello_APL_2016, Casalbuoni_NucInstr_2005}.

The dependence of $\lambda$ as a function of depth is then assumed to be:
\begin{equation}
    \lambda(x)=\left(\lambda_s-\lambda_0\right)\text{Erfc}\left[ \dfrac{x}{\delta}\right]+\lambda_0\text{,}
    \label{eq.lambda}
\end{equation}
where Erfc is the complementary error function, $\delta$ represents the diffusion length of the impurities and $\lambda_s$ the penetration depth value at the surface.

\begin{figure}[t]
\centering
\includegraphics[scale=1]{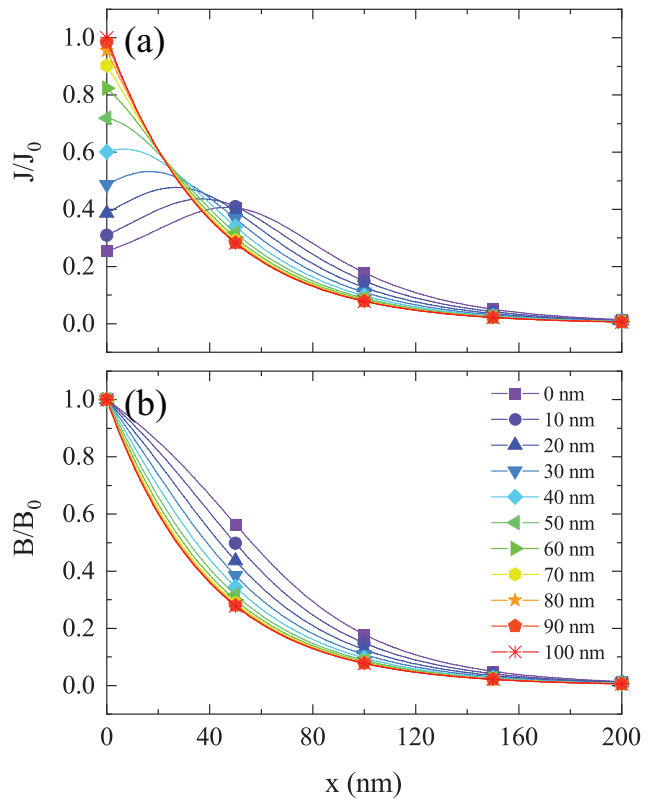}
\caption{Simulation of current redistribution for $\delta=50$~nm, $\lambda_s=100$~nm, and peak magnetic field for different values of material removal.}
\label{fig.simulation}
\end{figure}

Equation~\ref{eq.london} can be solved numerically imposing the magnetic field parallel to the surface at $x=0$ and defining the boundary conditions as $B(0)/B_0=1$ and $B'(\infty)/(\mu_0 J_0)=0$. Where $B_0$ and $J_0$ are respectively the peak magnetic field and the current density at the surface when no impurity profile is considered. This choice allows to observe how the current density varies in the presence of an impurity profile as a function of material removal, keeping the magnetic field value at the surface fixed. The current density is shown in Fig.~\ref{fig.simulation}(a), while the magnetic field profile is shown in Fig.~\ref{fig.simulation}(b).

The lambda profile considered in the simulations is reported in Eq.~\ref{eq.lambda}, with $\delta=50$~nm and $\lambda_s=100$~nm. Prior to any material removal, the current at the surface of the N-infused material is 60\% lower compared to $J_0$ and a pronounced peak appears around 50~nm, while the peak magnetic field remains unchanged. Notice how the magnetic field profile is not anymore exponential. Near the surface the slope is lower implying less effective screening due to the occurrence of a higher concentration of impurities, while in the bulk it increases following the decreasing of $\lambda$.

As soon as material is removed, the impurity profile is shortened and the concentration at the surface decreases, lowering $\lambda_s$. The current density peak shifts closer to the surface, the surface $J$ grows, and the magnetic field profile tends to an exponential decay. Once the material removed is at least equal to the diffusion length $\delta$, then $J(0)\simeq J_0$ and both magnetic field and current density profiles approach the exponential trend as a function of depth.

In summary, since the peak magnetic field represents the current density integrated within the material, it is independent of the current distribution. The same peak magnetic field at the surface is achievable by different current density profiles and $J(0)$ can be tuned to lower values by tweaking the impurities distribution in the material, as also shown in Ref.~[\onlinecite{Wave_PRR_2019}].

This phenomenon can explain why we observe the HFQS onset moving as a function of the material removal. If the source of HFQS is a dissipation mechanism localized at the near surface, then by means of an impurity profile the current density can be redistributed away from the surface and allow the dissipation to appear at higher peak magnetic fields.

The suspected dissipation mechanism localized at the near surface expected to generate HFQS is related to the occurrence of nano-hydrides generated upon cooling. It was observed by means of elastic recoil detection\cite{Romanenko_SUST_2011} that, in small concentrations, hydrogen in niobium preferentially sits in the first 10~nm from the surface even at room temperature. This observation implies that upon cooling niobium hydrides preferentially precipitate at the surface, in the first 10~nm of material, as demonstrated by cryogenic electron microscopy measurements.\cite{Trenikhina_JAP_2015}

Normal-conducting nano-hydrides at the cavity surface behaves as superconductors because they are proximity-coupled\cite{Orsay_PKM_1967} to the niobium matrix and can be described as SNS junctions with critical current density $J_H\sim1/\text{Sinh}(d)$, \cite{deGennes_RevModPhys_1964} above which they behave as normal resistors\textemdash $d$ is the hydride dimension. As described in Ref.~\onlinecite{Romanenko_SUST_2013}, nano-hydrides occur with a certain dimension distribution and the HFQS onset corresponds to the breakdown condition of the largest hydride present. Successive transition of smaller size hydrides are postulated to contribute to the Q-factor drop generating HFQS.

Statistically, the HFQS onset measured in EP cavities is $B_{EP}\simeq100$~mT,\cite{Padamsee_Book2,Ciovati_PRSTAB_2010} which corresponds roughly to a critical current density $J_H\simeq20\cdot10^{7}$~A/cm$^2$. By introducing an interstitial impurities profile, the current maximum is shifted towards the bulk and the hydrides critical current $J_H$ at the surface is achieved at higher peak magnetic field values, moving the HFQS onset accordingly and explaining why N-infusion mitigates HFQS effectively.

\begin{figure}[t]
\centering
\includegraphics[scale=1]{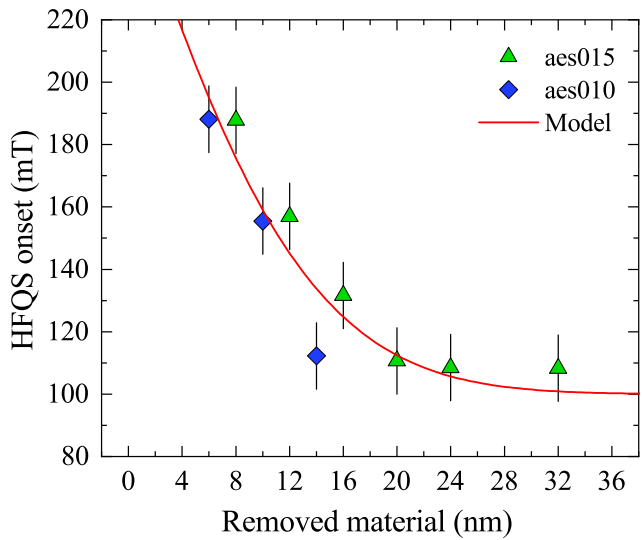}
\caption{HFQS onset data as a function of the material removal for the two cavities studied. The red solid line is the fit to the experimental data. Error in abscissa can be estimated \textit{a posteriori} as a cumulative error assuming 2~nm uncertainty per the material removal.}
\label{fig.fit}
\end{figure}

By removing material with subsequent HF rinses, the surface impurity content is lowered and the impurity diffusion profile shortened\textemdash the $\lambda$ profile follows accordingly. In turn, the current maximum shifts toward the surface and the hydrides critical current density at the surface is met at lower peak magnetic field values. We test this mechanism by fitting the HFQS onset data as a function of material removal shown in Fig.~\ref{fig.fit}, by means of Eq.~\ref{eq.london} and Eq.~\ref{eq.lambda}.

The HFQS onset field as a function of the material removed is defined as the field amplitude at which the Q-factor starts to drop. The error estimation was done \textit{a posteriori} assuming 25~\% on the peak magnetic field value.

Fixing the breakdown condition of hydrides at the surface by means of the boundary conditions $B'(0)=\mu_0J_H=-B_{EP}/\lambda_0$, with $B_{EP}=100$~mT, and imposing $B(\infty)=0$, we can solve Eq.~\ref{eq.london} for the peak magnetic field at which the breakdown condition is met and fit the data in Fig.~\ref{fig.fit}. The least square regression is carried out by leaving $\delta$ and $\lambda_s$ as free parameters, and considering the $\lambda$ profile of Eq.~\ref{eq.lambda} as being chopped at every material removal, returning $\delta=16.5\pm2.1$~nm and $\lambda_s=69.0\pm3.8$~nm. The result is shown in Fig.~\ref{fig.fit}. 

The fitting curve adheres nicely to the experimental data, suggesting that the HFQS onset variation as a function of material removal can be ascribed to the occurrence of an impurity profile with characteristic length scale of the order of ten to twenty nanometers. This result is in agreement with TOF-SIMS concentration profiles directly measured on N-infused cavity cut-outs prepared at $\SI{120}{\degreeCelsius}$,\cite{Romanenko_SRF_2019} where the diffusion profile of nitrogen was shown to have a length of the order of $\sim 15$~nm.

In the big picture, the proposed mechanism to prevent HFQS is expected to act in parallel with the mechanism for which hydrides precipitation is limited by the occurrence of vacancies\cite{Romanenko_APL_2013,Romanenko_APL_2014_1} and impurities.\cite{Grassellino_SUST_2013,Ford_SUST_2013,Garg_SUST_2018} As already stated above, nano-hydrides are forming at the cavity surface independently on the cavity processing.\cite{Sung_LCWS_2018,Sung_SRF_2019} The HFQS mitigation is then more complex: impurities and vacancies act as a deterrent for hydrides precipitation and the local $\lambda$ variation redistributes the current density away from the surface shifting the HFQS onset to higher peak magnetic fields.

Similarly to what we described in this Letter, the HFQS mitigation effect due to impurities profiles is expected to take place in \SI{120}{\degreeCelsius} baked cavities as well. LE-$\mu$SR measurements\cite{Romanenko_APL_2014_1} shows a non-exponential decrease of the magnetic field in \SI{120}{\degreeCelsius} baked cavities cut-outs which unequivocally indicates that currents are redistributed in the material. Differently from N-infusion, where this effect is most likely generated by interstitial nitrogen, for \SI{120}{\degreeCelsius} cavities the current redistribution is expected to be generated by an enriched layer of vacancies.\cite{Romanenko_APL_2014_1}

Concluding, in this Letter we demonstrated that the mitigation action against HFQS of N-infusion resides in the first tens of nanometers of material and that it is related to an interstitial impurities profile underneath the native oxide. Once the impurity profile is removed by a sequence of HF rinses the cavity behavior is reverted back to an EP-like behavior with HFQS onset around 100~mT. The model developed suggests that the impurity layer has a depth of about 15~nm which coincides with the nitrogen profile observed by means of TOF-SIMS in cavity cut-outs,\cite{Romanenko_SRF_2019} however other interstitial species\textemdash such as oxygen\textemdash cannot be ruled out.

The data that support the findings of this study are available from the corresponding author upon reasonable request.

We wish to thank A.~Romanenko for his valuable suggestions and advises. This work was supported by the United States Department of Energy, Office of High Energy Physics. Fermilab is operated by Fermi Research Alliance, LLC under Contract No. DE-AC02-07CH11359 with the United States Department of Energy.

\bibliography{Bibliography}
\end{document}